\documentclass[aps,pra,twocolumn,showpacs,final,nofootinbib,floatfix,reprint]{revtex4-1}

\usepackage{amsmath}
\usepackage{amsfonts}
\usepackage{amsthm}
\usepackage{amssymb}
\usepackage{graphicx}
\usepackage{braket}
\usepackage{setspace}
\usepackage{hyperref}
\usepackage{bbm}
\usepackage[english]{babel}
\usepackage{qcircuit}
\usepackage{tikz}
\usepackage{tikzscale}
\usepackage{standalone}
\usetikzlibrary{arrows}
\usetikzlibrary{positioning}
\usetikzlibrary{patterns}
\usepackage[normalem]{ulem}

\newtheorem{proposition?}{Proposition?}

\theoremstyle{definition}

\newcommand{\complex}{\mathbb C} 

\newcommand{\hi}{\mathcal{H}} 
\newcommand{\kb}[2]{|#1\rangle\langle #2|} 
\newcommand{\tr}[1]{\mathrm{tr}\left[#1\right]} 
\newcommand{\tre}[2]{\textrm{tr}_{#2}[#1]} 

\newcommand{\id}{\mathbb{I}} 

\newcommand{\E}{\mathsf{E}}

\newcommand{\Li}{\mathcal{L}}

\newcommand{\meas}{{\rm d}}

\newcommand{\sys}{\mathcal{S}}
\newcommand{\env}{\mathcal{E}}
\newcommand{\abs}[1]{\lvert #1\rvert}

\newcommand{\tOmega}{{\tilde\Omega}}
\newcommand{\tlambda}{{\tilde\lambda}}

\newcommand{\qwd}[1][-1]{\ar @{.} [0,#1]}
\newcommand{\ghostd}[1]{*+<1em,.9em>{\hphantom{#1}} \qwd}

\begin{document}

\title{Collision model for non-Markovian quantum dynamics}

\author{Silvan Kretschmer}
\affiliation{Institut f{\"u}r Theoretische Physik, Technische Universit{\"a}t Dresden, 
D-01062,Dresden, Germany}

\author{Kimmo Luoma}
\affiliation{Institut f{\"u}r Theoretische Physik, Technische Universit{\"a}t Dresden, 
D-01062,Dresden, Germany}

\author{Walter T. Strunz}
\affiliation{Institut f{\"u}r Theoretische Physik, Technische Universit{\"a}t Dresden, 
D-01062,Dresden, Germany}

\begin{abstract}
 We study the applicability of collisional models for non-Markovian dynamics of open 
 quantum systems.  
 By allowing interactions between the separate environmental degrees of freedom 
 in between collisions 
 we are able to construct a collision model that allows to study 
 quantum memory effects in open system dynamics. 
 We also discuss the possibility to embed non-Markovian collision model dynamics into 
 Markovian collision model dynamics in an extended state space.
 As a concrete example we show how using the proposed class of  
 collision models we can discretely model non-Markovian amplitude damping of a qubit.
 In the time-continuous limit, we obtain the well-known results
 for spontaneous decay of a two level system in to a structured zero-temperature reservoir. 
\end{abstract}


\maketitle

\section{Introduction}\label{sec:introduction}
An open quantum system consists of a system of interest ($\sys$, the open system) and an environment 
($\env$). Usually, the effects of the environment onto the open system dynamics 
are incorporated into an effective description in terms of master 
equations~\cite{alicki_quantum_1987,breuer_theory_2007} for the reduced density operator of the
open system. In general, the validity of the master 
equation approach is guaranteed in a Markovian setting, where the master equation takes the 
famous Gorini-Kossakowski-Sudarshan-Lindblad (GKSL) form 
and is a generator of a dynamical semigroup \cite{gorini_completely_1976,lindblad_generators_1976}.
Often, when starting from the microscopic model for $\sys+\env$, a series 
of approximations has to be made in order to be able to describe the open system dynamics 
in terms of the GKSL-master equation \cite{breuer_theory_2007,cohen-tannoudji_atom-photon_1998}. Naturally, 
these approximations
are not generally valid and the dynamical semigroup is not an exact description for all possible physical 
situations of interest~\cite{garraway_nonperturbative_1997}.

Collision models offer an alternative route to the description of open quantum system dynamics
\cite{rau_relaxation_1963,scarani_thermalizing_2002,ziman_diluting_2002,ziman_all_2005,ziman_description_2005,
giovannetti_dynamical_2005,
attal_repeated_2006,pellegrini_non-markovian_2009,giovannetti_master_2012-1,giovannetti_master_2012,
rybar_simulation_2012,ciccarello_collision-model-based_2013,caruso_quantum_2014,lorenzo_heat_2015}.
This description is appealing because the reduced system can be obtained in many cases without 
any approximations, hence the complete positivity (CP) of the dynamical map is guaranteed.
Collision models 
also provide a physically transparent way to introduce indirect measurements and conditional 
states for the open system \cite{attal_repeated_2006,pellegrini_non-markovian_2009}.
One can look at collision models as an approximation of continuous in time quantum dynamics but 
the study of discrete-in-time dynamics is interesting in its own right as we 
will show in this article.

A collision model consists of a system that interacts locally in time with different environmental
degrees of freedom \cite{ziman_diluting_2002}. 
As such, the description gives the dynamics discretely in time in terms of maps 
$\Phi_n:\sys(\hi_\sys)\to\sys(\hi_\sys)$, where $\sys(\hi_\sys)$ is the state space of the open system.
If, at each collision, the open system interacts with a new 
{\it uncorrelated} environmental degree of freedom then the  
dynamical map satisfies the semigroup property: $\Phi_{n+m}=\Phi_n\Phi_m$ \cite{scarani_thermalizing_2002,
ziman_diluting_2002,ziman_description_2005,ziman_all_2005,giovannetti_master_2012-1,giovannetti_master_2012},
see also Fig.~\ref{fig:MarkovCollisionModel}. Sometimes we call
separate environmental degree of freedoms sub-environments.

During recent years there has been considerable interest to generalize the collision model description
beyond Markovian quantum processes, ie. beyond the dynamical semigroup 
\cite{giovannetti_dynamical_2005,pellegrini_non-markovian_2009,rybar_simulation_2012,
ciccarello_collision-model-based_2013,caruso_quantum_2014,lorenzo_heat_2015}. 
One way to do this is to introduce correlated localized bath states \cite{rybar_simulation_2012,caruso_quantum_2014}.
For example by using 
correlated localized baths any indivisible quantum channel can be simulated \cite{rybar_simulation_2012}.
For our purposes the 
pre-correlated bath particles are not suitable because we want to introduce the 
memory as a part of the collision model dynamics.
A second way to introduce memory effects to the open system dynamics is to 
consider a system that sequentially collides with the same local environment but the 
evolution of the local environment is given by a quantum channel between the 
collisions. This type of model was considered in \cite{giovannetti_dynamical_2005}
for the purpose of processing quantum information.
A third, and the most relevant way for our purposes, is to allow an interaction
between the separate environmental degrees of freedom in between collisions. In \cite{ciccarello_collision-model-based_2013} the 
authors considered a partial swap interaction between the separate environmental degrees of 
freedom  that allows to
propagate correlations from earlier collisions forward in time. The authors were able to 
obtain a general form for the master equation in the continuous time limit. However,
a link to the exactly solvable model for spontaneous decay of a two 
level system into a structured reservoir \cite{garraway_nonperturbative_1997} is still missing and
will be established based on our novel collision model. Therefore,
our work is in close analogy of that in \cite{ciccarello_collision-model-based_2013} since we also want to
introduce tunable memory effects by incorporating interactions between the environmental degrees 
of freedom, but we take this interaction to be unitary. 

The structure of the paper is the following. In Sec. \ref{sec:model} we introduce the concept of 
a collision model in detail. First, we discuss a construction that leads always to a 
CP-divisible discrete dynamical map. Then we shift our focus to 
a generalized model that allows to study indivisible
quantum dynamical maps and non-Markovian dynamics.
Section  \ref{sec:discrete-maps} contains 
an application of the collision model, we show how time continuous amplitude damping
can be simulated using our discrete collision model. In Sec.~\ref{sec:non-mark-open} we discuss in detail
the divisibility and non-Markovianity properties of the example system.
In Sec.~\ref{sec:mark-embedd-non} we show how possibly non-Markovian collision model dynamics can be 
embedded into Markovian collision dynamics for an extended system 
and lastly in Sec.~\ref{sec:conclusions} we conclude.

\section{Collision models}\label{sec:model}
In a collision model the system of interest (the open system) interacts locally in time in a discrete
way with separate degrees of freedom of the environment.
A suitable Hilbert space for the collision model is $\hi=\hi_S\bigotimes_{k=1}^n\hi_{e,k}$, where
$\hi_S$ is the Hilbert space of the system and $\hi_{e,k}$ is the 
Hilbert space for $k$-th environment particle. There is in general no restriction on the 
possible initial state $\ket{\Psi_0}\in\hi$  of the collision model. However, we aim to connect the open system
dynamics of the collision model in the continuous limit to a dynamics given by a dynamical map 
$\Phi_t$ and therefore we assume that the initial state is of the form $\ket{\Psi_0}=\ket{\varphi}\otimes\ket{\omega}$.
Now, state $\ket{\omega}\in\bigotimes_k\hi_{e,k}$  could be generally correlated, and it 
is sometimes advantageous to consider correlated initial states \cite{rybar_simulation_2012} but we aim
in this work to model quantum memory effects dynamically and therefore we choose the 
initial state to be $\ket{\omega}=\bigotimes_k\ket{0_k}$. 
In general we focus on such collision models that might have local dynamics on the 
open system only, given by unitary operator $U_0$, interaction of the 
open system and $i$th sub-environment, given by unitary operator $W_i$, and 
possibly a unitary coupling between the sub-environments $i$ and $j$ given by $V_{ij}$.

Before going into details of the collision models we define for further discussion the 
concept of divisibility.  We call the dynamical map $\Phi_n\equiv\Phi_{n,0}$ {\it divisible}  
if the dynamical map can be composed 
as $\Phi_{n+m}=\Phi_{m+n,n}\Phi_{n}$, where $\Phi_{n+m,n}$ is a completely positive and trace preserving (CPT)
map. If the ``two times'' map $\Phi_{k,l}$ is 
not CPT for some pair $(k,l)$ then the dynamics is said to be {\it indivisible}. 
The two times map is defined as $\Phi_{n+m,n}=\Phi_{n+m}\circ\Phi_{n}^{-1}$ \cite{vacchini_markovianity_2011}.
These notions generalize to continuous families of maps in a trivial way.

\subsubsection{Collision model for divisible quantum dynamics}\label{sec:coll-model-divis}
\begin{figure}
  \begin{align*}
    \Qcircuit @C=0.5em @R=1.5em @!C{
    \lstick{\ket{\varphi}} & \gate{U_0} & \multigate{1}{W_1} & \gate{U_0}&\multigate{2}{W_2}&\qw\\
    \lstick{\ket{0}}       &\qw         &\ghost{W_1}         &\qw       &\ghostd{W_2}     &\qwd\\
    \lstick{\ket{0}}       & \qw        &\qw                 &\qw        &\ghost{W_2}      &\qw\\
    }
  \end{align*}
  \caption{\label{fig:MarkovCollisionModel} First two steps of a collision model where the open system, 
    initialized in state 
    $\ket{\varphi}$, interacts with a new and uncorrelated sub-environment at each collision. All
    sub-environments are initially at state $\ket{0}$.}
\end{figure}
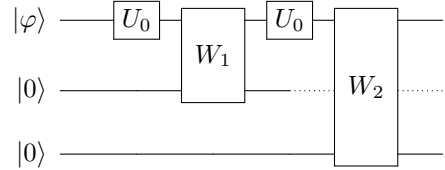

The simplest possible model fitting to our above described scenario emerges when  
we choose $V_{ij}=\id$. The first two steps of this type of model is presented in Fig.~\ref{fig:MarkovCollisionModel}.
The open system dynamics for this type of collisional model is given by
$\rho_0=\kb{\varphi}{\varphi}\mapsto \rho_n=\mathcal{E}_n(\rho_0)=
\tre{(W_n U_0 \cdots W_1 U_0)\rho_0\otimes\kb{\omega}{\omega}(U^\dagger_0 W_1^\dagger\cdots U^\dagger_0 W_n^\dagger)}{E}$. It is clear that 
the dynamical map $\mathcal{E}$ satisfies discrete semi-group property 
$\mathcal{E}_{n+m}=\mathcal{E}_n\mathcal{E}_m$, where $m,n\in \mathbb{Z}_+$. Also 
$\mathcal{E}_n=\mathcal{E}^n$ clearly holds.
The continuous limit of a collision model can be thought 
to emerge when $\mathcal{E}_t\approx \mathcal{E}_g^n$, where 
$n=[t/g^2]$ (where $[x]$ is a nearest integer to $x$) and $g$ is some parameter of the collision model 
ultimately to be related to the small time interval during which single collisions take place \cite{rybar_simulation_2012}.
A heuristic way to obtain the GKSL master equation, ie. the continuous limit for this collision model is to 
use the relation $\dot{\rho_t}=\frac{\meas}{\meas t}\mathcal{E}_t\mathcal{E}_t^{-1}(\rho_t)=\mathcal{L}\rho_t$
valid for continuous dynamical maps, whenever the inverse exists.
We would also like 
to point out the trivial observation that the  $(i-1)$th environment particle does not 
participate after the $(i-1)$th collision to the dynamics anymore.

As an example we consider a collision model where system and environment consists of 
qubits and initially all the environment qubits are set to their ground state $\ket{0}$.
We take the interaction between the system and the $i$th sub-environment  during 
a collision to be
\begin{align}
  W_i=&e^{-ig(\sigma_+ \otimes \sigma_-^{(i)} + \sigma_- \otimes \sigma_+^{(i)})},\label{eq:31}
\end{align}
where $g$ describes the coupling strength and we set 
$U_0=\id$ \cite{brun_simple_2002}. 
If we then set $g=\sqrt{\gamma \delta t}$,
take the derivative with respect to $\delta t$ and 
find the inverse map we can conclude after   
expanding the resulting expression to first order in $\delta t$ 
that in the limit 
$\delta t \to 0$ the collision model approximates the amplitude damping dynamical semigroup generated by
\begin{align}
  \mathcal{L}\rho=& \gamma \sigma_-\rho\sigma_+-\gamma\frac{1}{2}\{\sigma_+\sigma_-,\rho\}.
\end{align}

\subsubsection{Collision model for indivisible quantum dynamics}\label{sec:coll-model-indiv}
Our main goal is to describe indivisible quantum dynamics, using collisional models 
where we allow  for interaction between the separate environmental degrees of freedom. 
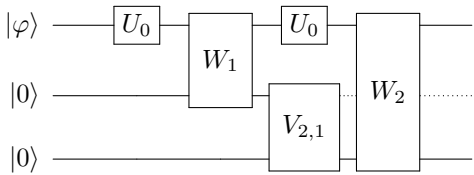
\begin{figure}
  \begin{align*}
    \Qcircuit @C=0.5em @R=1.5em @!C{
    \lstick{\ket{\varphi}} & \gate{U_0} & \multigate{1}{W_1} & \gate{U_0}&\multigate{2}{W_2}&\qw\\
    \lstick{\ket{0}}       &\qw         &\ghost{W_1}         &\multigate{1}{V_{2,1}}       &\ghostd{W_2}     &\qwd \\
    \lstick{\ket{0}}       & \qw        &\qw                 &\ghost{V_{2,1}}        &\ghost{W_2}      &\qw
                                                                                                         }
  \end{align*}
  \caption{\label{fig:NonMarkovCollision} First two steps of a collision model 
    where correlations are propagated by the interaction between the different sub-environments.
    Initial state of the system is $\ket{\varphi}$ and all environmental degrees of freedom  are 
    initially  at state $\ket{0}$.}
\end{figure}
The simplest possible way to do this to allow only nearest neighbor interactions. 
The dynamical map for the open system is 
\begin{align}\label{eq:33}
  \rho_n\equiv\Phi_n(\kb{\varphi}{\varphi})=&\tre{K_n\left(\kb{\varphi}{\varphi}\otimes\kb{\omega}{\omega}\right)K_n^\dagger}{E},
\end{align}
where $K_1= W_1 U_0$ and $K_n$ is defined iteratively as 
$K_n=W_nU_0V_{n,n-1}K_{n-1}$ for $n\geq 2$.
A schematic presentation of such model 
can be found in Fig.~\ref{fig:NonMarkovCollision}.
Clearly, the state of the 
$i$th sub-environment is generally not $\ket{0_i}$ when $W_i$ is applied because
before the $i$th collision the uncorrelated $i$th sub-environment 
first interacts with $(i-1)$th sub-environment via $V_{i,i-1}$ and after that 
the system interacts with $i$th environmental degree of freedom via $W_i$.  

The question of constructing indivisible quantum dynamics is thus a question of choosing suitable
two qubit unitary $V_{j+1,j}$. Obviously, if we choose  $V_{j+1,j}=U_A\otimes U_B$ the construction is 
essentially equivalent to the one presented in Fig.~\ref{fig:MarkovCollisionModel} since 
$U_A$ can be absorbed to $W_j$ and $U_B$ is just a rotation of the initial state.
Therefore the unitary operator $V_{i+1,i}$ has to be at least non-separable. 

Intuitively it is clear that by introducing the coupling $V_{i+1,i}$ we can propagate information 
from the earlier collisions trough the correlations created
between $\hi_{e,i+1}\otimes\hi_{e,i}$
and these correlations might  have influence on the nature of the dynamics. 
To verify this intuition we first tested the collision model construction by 
sampling  pairs of two qubit unitaries $Z_1$ and $Z_2$ from the uniform Haar measure \cite{mezzadri_how_2007} 
and using those as 
a building block for the collision model. Specifically we constructed 
the maps $\Phi_2$,$\Phi_1$ and $\Phi_{2,1}=\Phi_2\circ\Phi_1^{-1}$ according to Eq.~\eqref{eq:33}. By sampling $10^6$ pairs $(Z_1,Z_2)$ 
we observed that 
approximately $17\%$ of the maps $\Phi_2$ were indivisible, ie. $\Phi_{2,1}$ was not 
completely positive. 

As a second test we chose randomly pairs of two qubit unitaries $V=e^{-i\sum_{j=1}^3 \alpha_j\sigma_j\otimes\sigma_j}$ such that 
$\pi/4\geq\alpha_1\geq\alpha_2\geq\alpha_3\geq 0$,
$\alpha_1+\alpha_2\geq \pi/4$ and $\alpha_2+\alpha_3\leq \pi/4$. These correspond to maximally entangling unitaries \cite{kraus_optimal_2001}.
In this case by sampling $10^6$ pairs we obtained that  $35\%$ of samples resulted in indivisible dynamical maps.
From these two tests we can conclude that the proposed collision model (see Eq.~\eqref{eq:33} and 
Fig.~\ref{fig:NonMarkovCollision}) is suitable for the study of general open system dynamics.

\subsubsection{Collision models and repeated measurements}\label{sec:coll-models-repe}
Further, we would also like to point out  
that the collision model construction presented in Fig.~\ref{fig:NonMarkovCollision} 
can be used for studying repeated quantum measurements \cite{brun_simple_2002,attal_repeated_2006,pellegrini_non-markovian_2009} 
in a setting where the average
dynamics does not form a semigroup.
Due to only having a nearest neighbor interaction between the sub-environments
the $(i-1)$th environmental degree of freedom  does not participate after the $i$th collision to the dynamics 
anymore. In Figs.~\ref{fig:MarkovCollisionModel} and \ref{fig:NonMarkovCollision} 
the non-participating environmental degrees of freedom are indicated with a dashed line. Therefore 
after the $i$th collision the state of the sub-environments  $k\leq i$ can be measured without 
altering the future evolution. This happens because any observable $\E_x^{(k)}=(M_x^{(k)})^\dagger M_x^{(k)}$ 
on $\hi_{e,k}$ commutes with $W_i$, for $i\neq k$ and $V_{i,i-1}$ for $k\neq i \wedge k\neq i-1$. To state this idea as simply as 
possible we refer to Fig.~\ref{fig:NonMarkovCollision}  and to the total state after just two steps. 
Let $U_0=\mathbb{I}$, then the total state {\it with measurement} of the first sub-environment right 
after interaction $V_{2,1}$ is 
$\varrho_T =  \sum_{x_1} W_2 M_{x_1}^{(1)}V_{2,1}W_1\rho_0\otimes\kb{\omega}{\omega} W_1^\dagger V_{2,1}^\dagger M_{x_1}^{(1)}W_2^\dagger$.  
Let $\rho_T = K_2\rho_0\otimes\kb{\omega}{\omega}K_2^\dagger$ ( with $U_0=\mathbb{I}$).
Since $[W_2,M_{x_1}^{(1)}]=0$ it is clear that both $\rho_T$ and $\varrho_T$ lead to the same reduced state $\rho_2$.
By continuing this way and adding a measurement on $\hi_{e,k-1}$ after $V_{k,k-1}$ in the collision 
model we could repeatedly measure the collision model trajectories without altering the 
mean dynamics. It should be noted that by this procedure we obtain 
trajectories for two qubit states, namely a repeated measurement up to 
step $k$ results in a conditional state on $\hi_S\otimes\hi_{e,k}$.

After discussing the potential generality of the collision model approach on open system dynamics 
we will focus on using a collision model approach to simulate continuous in time 
dynamics trough a paradigmatic example. 

\section{Collision model description for time continuous amplitude damping}\label{sec:discrete-maps}
The process of spontaneous emission of a two level system to a structured reservoir 
can be described with the following exact master equation (in a suitable interaction picture) 
\begin{align}\label{eq:34}
  \dot \rho_t =&-\frac{i}{2}s_t[\sigma_+\sigma_-,\rho_t]+\gamma_t\left(\sigma_-\rho\sigma_+-\frac{1}{2}\{\sigma_+\sigma_-,\rho\}\right),
\end{align}
where $s_t$ is a time dependent shift on the transition energy of the two level system and  
$\gamma_t$ is a time dependent decay rate which may be temporarily negative. Functions $s_t$ and $\gamma_t$ are  
related to the zero temperature bath correlation function $\chi_{t,s}$ \cite{garraway_nonperturbative_1997,strunz_open_1999}.
The solution of the master equation can be given as 
\begin{align}\label{eq:35}
  \rho_t \equiv& \Phi_t^{(c)}\rho=\rho_{11}|\eta_t|^2\kb{1}{1}+
                         (1-\rho_{11}|\eta_t|^2)\kb{0}{0}\notag\\
  &+\rho_{10}\eta_t\kb{1}{0}+\rho_{01}\eta^*_t\kb{0}{1},
\end{align}
where $\abs{\eta_t}^2 \in [0,1]$. The time dependent function 
$\eta_t$  
satisfies the following differential equation 
$\dot\eta_t = -\lambda\int_0^t\meas s\,\chi_{t-s}\eta_s$ and  
initial condition $\eta_0=1$. $\lambda$ is an additional parameter for 
the coupling strength between system and the environment.
The resulting dynamical map 
$\Phi_t$ describes amplitude damping with a time dependent parameter $\abs{\eta_t}^2$ \cite{nielsen_quantum_2004}. 

The two times map $\Phi^{(c)}_{s+t,t}=\Phi^{(c)}_{t+s}\circ(\Phi^{(c)}_t)^{-1}$ can be written as 
\begin{align}\label{eq:36}
  \Phi^{(c)}_{t+s,t}\rho=&\rho_{11}|\frac{\eta_{t+s}}{\eta_t}|^2\kb{1}{1}+
                         (1-\rho_{11}|\frac{\eta_{t+s}}{\eta_t}|^2)\kb{0}{0}\notag\\
                   &+\rho_{10}\frac{\eta_{t+s}}{\eta_t}\kb{1}{0}+\rho_{01}\frac{\eta^*_{t+s}}{\eta_t^*}\kb{0}{1}. 
\end{align}
One sees that this is again an amplitude damping channel iff $\abs{\frac{\eta_{t+s}}{\eta_t}}^2\in [0,1]$.
We denote by $M_{t+s,t}$ the Choi matrix of the map $\Phi_{t+s,t}$ \cite{choi_completely_1975}. 
From $M_{t+s,t}$ we see that $\Phi_{t+s,t}$ is  completely positive when 
$\abs{\eta_{t+s}}^2 \leq \abs{\eta_t}^2$. 
Thus the dynamical map is indivisible whenever the above condition is violated. 
This corresponds to a temporarily negative decay rate $\gamma_t$ in 
the master equation \eqref{eq:34} which corresponds to $\frac{\meas}{\meas t}\abs{\eta_t}^2>0$ 
\cite{laine_measure_2010}.

To fully specify the considered model, we take the spectral density of the 
reservoir to be a single Lorentzian leading to
an exponential bath correlation function 
$\chi_t=\frac{\Gamma}{2}e^{-\Gamma t-i\Omega t},\,t\geq 0$. 
From now on, with the exception of Sec. \ref{sec:cont-limit-lambd},  we will use dimensionless units 
\begin{align}
  \Gamma t=\tau,\,\, \tOmega=\Omega/\Gamma,\,\, \tlambda=\lambda/\Gamma,\label{eq:43} 
\end{align}
which allow us to write 
\begin{align}
  \dot\eta_\tau=&-\frac{\tlambda}{2}\int_0^\tau\meas\tau'e^{-(\tau-\tau')-i\tOmega(\tau-\tau')}\eta_{\tau'}.\label{eq:45}
\end{align}
The well known analytical solution in these units is
\begin{align}\label{eq:19}
  \eta_\tau =& e^{-\tau-i\tOmega\tau}\left(\frac{1+i\tOmega}{b}\sinh\left(\frac{\tau}{2}b\right)+
               \cosh\left(\frac{\tau}{2}b\right)\right),
\end{align}
where $b=\sqrt{(1+i\tOmega)^2-2\tlambda}$. 

To use collision model \eqref{eq:33} to simulate discontinuously the map 
\eqref{eq:35} we require that at the $n$th step
the discrete  collision model dynamics approximates the time-continuous 
dynamics given by Eq.~\eqref{eq:35}, ie. 
$\Phi_n\approx \Phi_{\tau=n\delta \tau}^{(c)}$.
First of all, the environment is taken to consist of qubits, ie. $\hi_{e,i}=\complex_2$, and all
environment qubits are initially prepared to be in their ground state, $\ket{0_i}$.
As before, we take the interaction between the system and environment qubits 
to be $W_i$, see Eq.~\eqref{eq:31}. We also choose the interaction
between the environment qubits to be similar to $W_i$ but 
with independent parametrization
\begin{align}\label{eq:2}
  V_{n,n-1}=&e^{-iG(e^{i\phi}\sigma_+^{(n-1)}\otimes\sigma_-^{(n)}+e^{-i\phi}\sigma_-^{(n-1)}\otimes\sigma_+^{(n)})},
\end{align}
where $G$ is a real valued coupling constant and  $\phi$ is an arbitrary but 
fixed phase which is to be determined later.
Lastly, we choose $U_0=\mathbb{I}$. The collision
model is conveniently parametrized with parameters $(g,G,\phi)$.

The excitation number operator for the collision model is 
$N=\sigma_+\sigma_- + \sum_k\sigma_+^{(k)}\sigma_-^{(k)}$.
Since $[V_{n,n-1},N]=[W_i,N]=0$, the excitation number is a conserved quantity.
From the initial condition 
$\kb{0}{0}$ for the environment qubits and from the excitation number conservation follows
that there can be only one excitation in the collision model if the initial state of the 
system qubit contains it.
We set $\rho$ to be the initial state of the system qubit. By plugging in $V_{i,i-1}$, $W_i$ and $U_0$ to Eq.~\eqref{eq:33} we find 
that the state of the system qubit, $\rho_n$ after $n$ collisions is
\begin{align}
  \rho_n=&\Phi_n(\rho)=\rho_{11}|\eta_n|^2\kb{1}{1}+
                         (1-\rho_{11}|\eta_n|^2)\kb{0}{0}\notag\\
  &+\rho_{10}\eta_n\kb{1}{0}+\rho_{01}\eta^*_n\kb{0}{1},\label{eq:44}
\end{align}
which shows that the collision model dynamics for the system qubit is of amplitude damping type.
The discrete function $\eta_n$ satisfies the following equation
\begin{align}\label{eq:32}
  \eta_n=& \cos g\bigg(\eta_{n-1}-\tan^2g\sum_{j=0}^{n-2}\eta_j(-ie^{i\phi}\cos g\sin G)^{n-1-j}\bigg),
\end{align} 
see Sec.~\ref{sec:mark-embedd-non} for the derivation. $\eta_n$ depends on the 
past values $\eta_j$, $j\leq n-2$, through a discrete memory kernel which 
emerges from the ``single step memory'' $V_{i,i-1}$ in the collision
model that propagates information from the past collisions. 

\begin{figure}
  \includegraphics{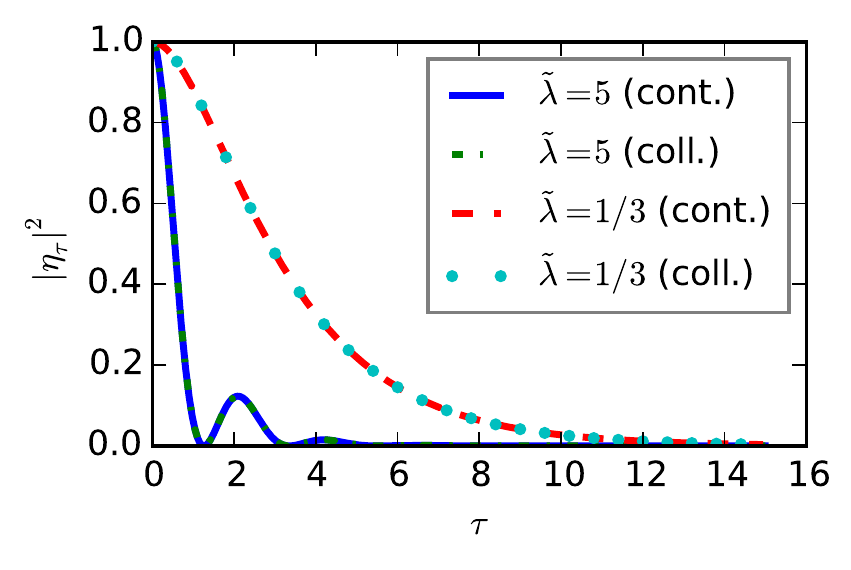}
  \caption{\label{fig:eta} (Color online) Comparison of $\eta_\tau$ obtained from 
    continuous and collision model. 
    Case $\tlambda=1/3$ corresponds to Markovian evolution and 
    $\tlambda=5$ to non-Markovian evolution of the open system.
    Other parameters are  
    $\delta \tau = 0.02\tlambda^{-1}$ and $\tOmega=0$. Collision model 
    parameters $G,g$ and $\phi$ are obtained using Eqs.~\eqref{eq:22}-\eqref{eq:30}.}
\end{figure}

The only question that is left is if the time-continuous limit of the 
collision model dynamics 
given by Eqs.~\eqref{eq:33},\eqref{eq:44}, and \eqref{eq:32}  corresponds to 
the time continuous amplitude damping dynamics of Eqs.~\eqref{eq:35},\eqref{eq:19}.
Clearly, if $\eta_n$ coincides with $\eta_\tau$ in some limit the answer is 
affirmative.
We find that by choosing the parameters $(g,G,\phi)$ of the collision model 
as 
\begin{align}
  G =& \arcsin(e^{-\delta \tau}),\label{eq:22}\\
  \phi =& \pi/2-\tOmega\delta \tau,\label{eq:26}\\
  g =& \sqrt{\frac{\tlambda}{2}}\delta \tau,\label{eq:30}
\end{align} 
where $\delta \tau$ is the dimensionless time interval between two consecutive collisions, 
that the difference $\abs{\eta_n-\eta_\tau}\to 0$ in the limit $\delta\tau\to 0$.
We verify our results in Fig.~\ref{fig:eta}. 
This choice of relation between parameters $(g,G,\phi)$ and $(\delta\tau,\tlambda,\tOmega)$ 
emerges for fixed $\tlambda$ by assuming that 
$g\propto\delta\tau$ and then demanding that $\frac{\eta_{n+1}-\eta_{n}}{\delta\tau}$ converges
to Eq.~\eqref{eq:45} in the limit $\delta\tau\to 0$.
For the proof see Appendix ~\ref{sec:proof-eqs.-eqref}.

\section{Comparison of time continuous and collision model dynamics}\label{sec:non-mark-open}
In this section we compare the continuous in time and collision model with each other in the 
case of amplitude damping dynamics for the system qubit. 
We discuss their differences in terms of divisibility or non-Markovianity. For simplicity 
we study only the resonant case, ie. we set $\tOmega=0$  
for the rest of this section. From Eq.~\eqref{eq:26} follows that $\phi=\pi/2$. 

Amplitude damping in the collision model is determined by the function $\eta_n$. 
From Eq.~\eqref{eq:32} we see that we reach all possible values of the function $\eta_n$
when $g,G\in [0,2\pi)$. The absolute value $\abs{\eta_n}$ is symmetric with respect to 
reflections of both $g,G$ with respect to $\pi/2$. Therefore we define the domain $S_\epsilon$ 
of the collision model parameters to be  
\begin{align}
  (g,G) \in S_\epsilon = [0+\epsilon,\pi/2-\epsilon)\times[0+\epsilon,\pi/2-\epsilon)\label{eq:46},
\end{align}
where $\epsilon\ll 1$.
Equations \eqref{eq:22}-\eqref{eq:30} give the transformation between the
parameters  $(\tOmega,\tlambda,\delta\tau)$ and $(G,g,\phi)$. 
The inverse transformation is 
\begin{align}
  \delta \tau =& \ln(1/\sin G),\label{eq:39}\\
  \tlambda =& \frac{2g^2}{\ln(1/\sin G)^2}.\label{eq:40}
\end{align}
The inverse function theorem states that the 
transformation $(g,G,\phi) \mapsto (\delta\tau,\tlambda,\tOmega)$ is invertible apart from some neighborhood of the 
singular point $\delta\tau=0$. 
Since the parameters of the collision model and the time continuous model are 
in one-to-one correspondence, we can re-parametrize $\eta_n$  
in terms of $\tlambda,\delta \tau$ and identify time step $n$ with a time $n\delta \tau = \tau$ in 
the time continuous model. We can thus write 
$\eta_n\equiv \eta_{n}^{(g,G)}=\eta_n^{(\delta \tau,\tlambda)}\equiv\eta_{n\delta \tau}=\eta_\tau$. 

\subsection{{Divisibility}}\label{sec:divisibility}
Open system dynamics 
is called indivisible in the time-continuous case if 
the two times map $\Phi^{(c)}_{\tau+\tau',\tau}$ is not completely positive, where 
$\tau,\tau'>0$. 
In the time-discrete case the definition is that the map is indivisible 
if $\Phi_{m+n,n}$ is not completely positive. Both maps are indivisible 
if the associated Choi matrix is not positive. We denote 
the time-continuous and the time-discrete Choi matrices with 
$M_{\tau+\tau',\tau}$ and $M_{m+n,n}$, respectively. 

The negativity condition of the Choi matrices $M_{n+1,n}$ in the time-discrete 
and $M_{\tau+\delta\tau,\tau}$ in the   
time-continuous cases is 
\begin{align}
  \abs{\eta_{n+1}}^2 >& \abs{\eta_n}^2,\label{eq:47}\\
  \abs{\eta_{\tau+\delta\tau}}^2 >&\abs{\eta_\tau}^2.\label{eq:48}
\end{align}
These are the conditions for the non-complete positivity  of the associated two times maps.
We know that the time-continuous model is indivisible when $\tlambda>1/2$ \cite{laine_measure_2010}.
In Fig. \ref{fig:nm_g_G_n30} we have plotted the boundary $\tlambda= 1/2$ on set $S_{\epsilon=0.01}$ with a solid gray line. 
Above this 
line the collision model dynamics is indivisible. In Fig. \ref{fig:nm_dt_Gamma_n30} we  have plotted the image 
of $S_{\epsilon=0.01}$ under the mapping \eqref{eq:39}\eqref{eq:40} with dark gray non-solid lines. 
The dotted dark gray line is the image of the boundary $(g,G)\vert_{g=\epsilon}$. Solid gray line is the boundary
$\tlambda=1/2$ between the divisible and indivisible dynamics. Light gray non-solid lines are the 
boundaries of the set $S_{\epsilon=0.005}$. One sees how the continuous limit $\delta\tau\to 0$ is obtained 
by letting $g\to 0$ for a fixed $\tlambda$. Colored data points in both figures correspond to 
non-Markovianity of the dynamics for $n=30$ collisions, which we will discuss next.  

\subsection{Non-Markovianity}\label{sec:non-markovianity}
Various different criteria for non-Markovian open system dynamics have been 
proposed in the literature recently, \cite{wolf_assessing_2008,breuer_measure_2009,rivas_entanglement_2010,lu_quantum_2010,
luo_quantifying_2012,lorenzo_geometrical_2013,giovannetti_master_2012,chruscinski_degree_2014}. For further 
detail see a recent review article \cite{breuer_non-markovian_2015}. In this work we use the 
measure proposed in \cite{breuer_measure_2009} which is based on contractivity of the trace distance 
under positive maps \cite{perez-garcia_contractivity_2006}. 
For an application of this non-Markovianity measure to time-discrete dynamics, see for example~\cite{luoma_discrete_2015}.

Let
$D=\frac{1}{2}\abs{\rho^{(1)}-\rho^{(2)}}$ be the trace distance between two quantum states.
In the time continuous case the measure for non-Markovianity is given as an optimization over initial state pairs
that maximize the time intervals of trace distance increase during some evolution interval $\tau'\in[\tau_0,\tau]$. It can be 
expressed as  
\begin{align}\label{eq:1}
  \mathcal{N}_\tau = \max_{\{\rho^{(i)}\}} \int_{\dot{D}(\tau')\geq 0} \dot{D}(\tau')\meas \tau'.
\end{align}
It has been shown that the optimal state pair is
orthogonal \cite{wismann_optimal_2012}.
Crucial difference between time-continuous dynamics and collision model dynamics is  that 
in the latter case the derivative of the trace distance does not exist. Therefore 
the measure in the time-discrete setting for $k\in[0,n]$ is given by
\begin{align}\label{eq:38}
  \mathcal{N}_n=& \max_{\{\rho^{(i)}\}}\sum_{D(k+1)-D(k)>0} D(k+1)-D(k).
\end{align}

\begin{figure}
\includegraphics{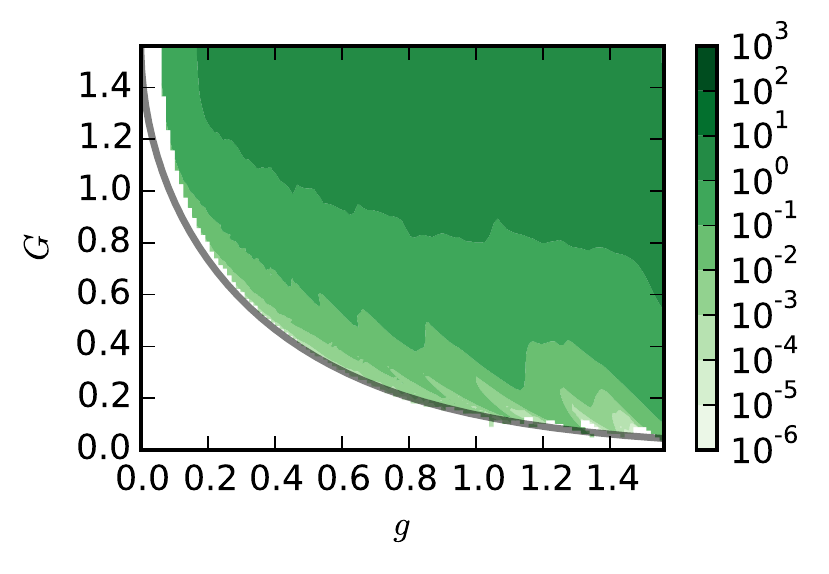}
\caption{\label{fig:nm_g_G_n30} (Color online) Non-Markovianity of the collision model after $n=30$ steps and using
the optimal initial state pair $\ket{\varphi_{\pm}}=\frac{1}{\sqrt{2}}\left(\ket{0}\pm\ket{1}\right)$.
We chose uniformly 14 400 values $g,G \in S_{\epsilon}$,
where $\epsilon=0.01$. We have chosen $\epsilon\neq 0$ in order to avoid divergences.
The black line corresponds to the boundary between divisible and indivisible 
dynamics of the continuous in time model, $\tlambda=1/2$.}
\end{figure}

\begin{figure}
\includegraphics{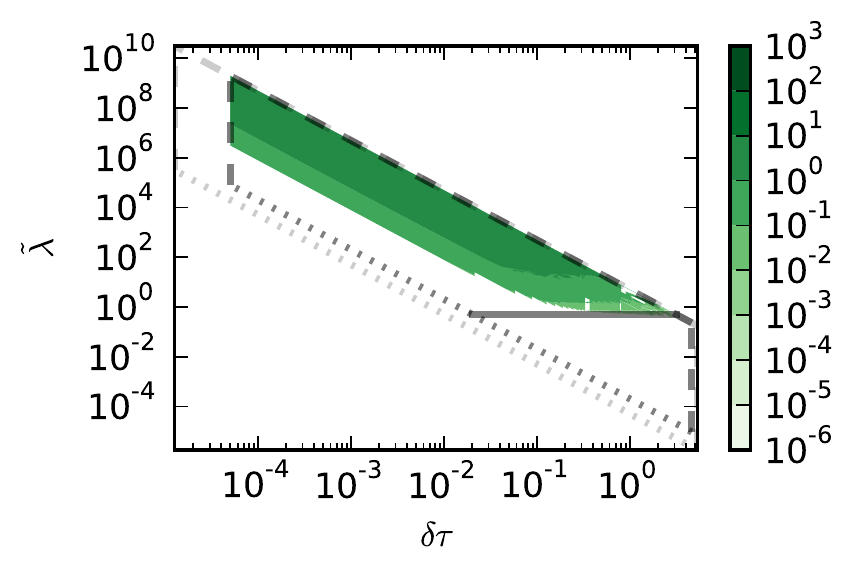}
\caption{\label{fig:nm_dt_Gamma_n30}
(Color online) Data from Fig.~\ref{fig:nm_g_G_n30} is mapped to 
$(\delta\tau,\tlambda)$-coordinates using Eqs.~\eqref{eq:39}, \eqref{eq:40}. 
Solid line is the boundary $\tlambda=1/2$ between divisible and indivisible 
dynamics.
Area confined by dark gray (non-solid) lines 
correspond to the image of  $S_\epsilon$,
where $\epsilon = 0.01$ under the same mapping. Light gray (non-solid) lines indicate
the image of $S_{\epsilon'}$ with $\epsilon' = 0.005$. 
The dotted lines correspond to the image of the boundary $g=\epsilon$ of the 
set $S_{\epsilon}$. We thus see how the time continuous limit $\delta \tau \to 0$ is 
obtained for fixed $\tlambda$ by letting $g\to 0$.}
\end{figure}

For the special case studied here 
we can express the trace distance between two initial states $\rho^{(1)}, \rho^{(2)}$ under either discrete or
continuous amplitude damping as
\begin{align}\label{eq:27}
  D(n)=&\frac{1}{2}\sqrt{\abs{\eta_n}^2(d_{x}^2+d_{y}^2)+\abs{\eta_n}^4 d_{z}^2},
\end{align}
where $d_{\alpha,n}=\tr{\sigma_\alpha (\rho^{(1)}-\rho^{(2)})}$, $\sigma_\alpha,\,\alpha=1,2,3$ are the 
Pauli matrices. 
The time-continues case is obtained by replacing the discrete index $n$ with a 
continuous parameter $\tau$. We know that for this case the 
optimal pair of initial states is any pair of states on the equator of the Bloch sphere 
\cite{xu_proposed_2010}. We choose the initial state pair to be $\ket{\varphi_{\pm}}=\frac{1}{\sqrt{2}}(\ket{0}\pm\ket{1})$
in order to obtain simple expression for the trace distance evolution 
between the optimal pair; $D^*(n)=\abs{\eta_n}$ in the
discrete case and $D^*(\tau)=\abs{\eta_\tau}$ in the time continuous case.
Since $D^*(\tau+\delta \tau)-D^*(\tau) = \abs{\eta_{\tau+\delta \tau}}-\abs{\eta_{\delta \tau}}$ 
we can deduce that whenever $\abs{\eta_\tau}$ behaves non-monotonically between
subsequent steps it is a signature of non-Markovianity of the dynamics. 
This observation explains why the definition $S_\epsilon$ in Eq.~\eqref{eq:46} is reasonable.
By recalling Eq.~\eqref{eq:47} we see that for this specific system the indivisibility of the 
two times map $\Phi_{n+1,n}$ implies the non-Markovianity of the dynamics. This is seen also in 
Figs.~\ref{fig:nm_g_G_n30} and \ref{fig:nm_dt_Gamma_n30}, where non-Markovianity occurs in the 
regions of indivisibility. 

In Fig.~\ref{fig:nm_g_G_n30} we chose 14 400  pairs of values 
for $(g,G) \in S_\epsilon$, where 
$\epsilon = 0.01$ and calculated the 
non-Markovianity of the collision model dynamics using the optimal initial pair for $n=30$ 
collisions. In this figure we
also plotted the boundary between divisible and indivisible dynamics, 
$\tlambda =1/2$ of the continuous in time model. We see that the non-Markovianity 
of the collision model dynamics occurs in the region of indivisibility
of the time-continuous model. Small inconsistencies are due to numerical inaccuracies.
Nonzero value for $\epsilon$ is chosen to avoid divergences. From the figure we see that 
by increasing the value of $G$, the strength of the environment particle coupling $V_{i,i-1}$, the 
dynamics becomes more non-Markovian. 

In Fig.~\ref{fig:nm_dt_Gamma_n30} we mapped the non-Markovianity values to $(\delta \tau,\tlambda)$ coordinates using 
Eqs.~\eqref{eq:39} and \eqref{eq:40}. We also mapped the boundaries of the region $S_\epsilon$ for
$\epsilon = 0.01$ with dark gray non-solid lines and for $\epsilon=0.005$ with
light-gray non-solid lines. Solid gray line is the boundary $\tlambda=1/2$. 
There are regions the image of $S_{\epsilon=0.01}$ such that 
$\tlambda>1/2$ that do not show non-Markovianity. These emerge because the number 
of collisions  $n=30$ is too small for the first revival of $\abs{\eta_{n\delta\tau}}$ to occur.   

Interestingly, in the case of amplitude damping dynamics, we can connect sub-optimal non-Markovianity of 
the collision model dynamics to the indivisibility of the map $\Phi^{(c)}_{\tau+\delta\tau,\tau}$ of 
the time-continuous case. It is based on the following observations.
The only possibly negative eigenvalue of the 
Choi matrix $M_{\tau+\delta\tau,\tau}$ is $\lambda_{\tau+\delta \tau,\tau}^{(M)} =\abs{\eta_\tau}^2-\abs{\eta_{\tau+\delta \tau,\tau}}^2$ 
Evolution of the trace distance in the collision model for the sub-optimal initial state pair $\ket{0},\ket{1}$ 
($d_z=2, d_x=d_y=0$ in Eq.~\eqref{eq:27}) 
is $\tilde D(k+1)-\tilde D(k)=\abs{\eta_{k+1}}^2-\abs{\eta_k}^2$. This just the possibly 
negative eigenvalue of the Choi matrix $M_{k+1,k}$.
Now, by Eqs.~\eqref{eq:39},\eqref{eq:40} and the discussion
in the beginning of the Sec.~\ref{sec:non-mark-open} we can write  
\begin{align}
  \tilde D(k+1)-\tilde D(k)&=\abs{\eta_{\tau+\delta\tau}}^2-\abs{\eta_\tau}^2=\lambda^{(M)}_{\tau+\delta\tau,\tau}.\label{eq:49}
\end{align}
By defining the sub-optimal non-Markovianity as usual 
$\mathcal{M}_n=\sum_{\tilde D(k)}\tilde D(k)$, we can write
\begin{align}\label{eq:41}
  \mathcal{M}_n =& -\sum_{\lambda_{\tau'+\delta \tau,\tau'}^{(M)}<0}\lambda_{\tau'+\delta \tau,\tau'}^{(M)}.
\end{align}
We would like to point out that $\mathcal{M}_n>0\implies\mathcal{N}_n>0$. We have thus shown that 
in this case the discrete non-Markovian dynamics provides us information about the 
indivisibility properties of the time-continuous dynamical map.

\section{Markovian embedding of non-Markovian dynamics}\label{sec:mark-embedd-non}
It is well known in the classical case that non-Markovian processes can be embedded into Markovian 
processes in an extended state space. This can also be done in the quantum case, a prime example 
being the pseudomode method \cite{garraway_nonperturbative_1997} where the indivisible dynamics 
of the system of interest is obtained by partial trace from a state of a larger system 
that obeys the GKSL master equation. In this section we study the possibility to embed the 
class of collisional models presented schematically in Fig.~\ref{fig:NonMarkovCollision} when
the open system and the environment consists of qubits.

By taking a partial trace over the Hilbert space $\bigotimes_{i=1}^{n-1}\hi_{e,i}$ 
of the total state $K_n(\kb{\varphi}{\varphi}\otimes\kb{\omega}{\omega})K_n^\dagger$ after $n$ collisions
we obtain a two qubit state 
$\kappa_n\in \mathcal{S}(\hi_\sys\otimes\hi_{e,n})$. The reduced state of the system qubit 
can be obtained from state $\kappa_n$ by tracing out the $n$th environment qubit
\begin{align}
  \rho_n=&{\rm tr}_n\left\{\kappa_n\right\}.\label{eq:3}
\end{align}
It turns out that 
the two qubit state then evolves according to the map $\Lambda$ defined by
\begin{align}
\kappa_n=&\Lambda\kappa_n\notag\\
  &\equiv{\rm tr}_{n-1}\left\{W_nV_{n,n-1}\left(\kappa_{n-1}\otimes\kb{0_n}{0_n}\right)V_{n,n-1}^\dagger W_n^\dagger\right\},\label{eq:6}
\end{align}
with an initial condition
\begin{align}  
  \kappa_1=&W_1\rho_0\otimes\kb{0_1}{0_1}W_1^\dagger.\label{eq:7}
\end{align}
The map $\Lambda$ does not depend on the step numbers since we assume that the collision 
model is homogeneous. Then clearly the state $\kappa_n$ can be written 
as $\kappa_n=\Lambda^{n-1}\kappa_1$, where $\Lambda^k=\Lambda\circ\Lambda^{k-1}$.
The structure of the collision model however gives a very special form for the initial 
state $\kappa_1$ of the enlarged dynamics, especially it might be an entangled state 
between the system and the first environment qubit.

From this follows our first non-trivial observation. The map $\Lambda$ is not necessarily completely
positive since it is obtained from tracing over $(n-1)$th qubit which might be 
entangled with the system qubit. However it is easy to see that the $\kappa_n$ is a valid state since 
the map $\Lambda$ is positive and trace preserving. Thus we have at least a model for 
positively divisible dynamics.

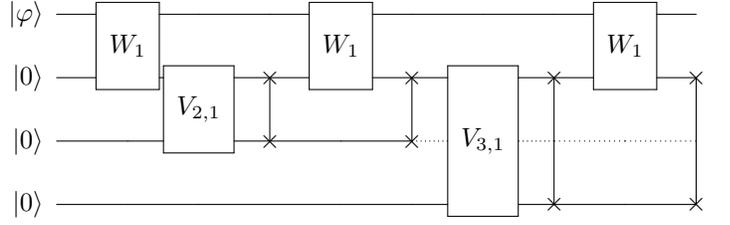
\begin{figure}
  \begin{align*}
    \Qcircuit @C=0.02em @R=1.5em @!C{
    \lstick{\ket{\varphi}} & \multigate{1}{W_1} & \qw                 &\qw          &\multigate{1}{W_1}&\qw &\qw                 &\qw        &\multigate{1}{W_1}&\qw\\
    \lstick{\ket{0}}       &\ghost{W_1}         &\multigate{1}{V_{2,1}}&\qswap       &\ghost{W_2}        &\qswap&\multigate{2}{V_{3,1}}&\qswap     &\ghost{W_3}    &\qswap\\
    \lstick{\ket{0}}       &\qw                 &\ghost{V_{2,1}}       &\qswap\qwx   &\qw              &\qswap\qwx &\ghostd{V_{3,1}}  &\qwd\qwx&\qwd       &\qwd\qwx\\
    \lstick{\ket{0}}       &\qw                 &\qw                  &\qw          &\qw              &\qw &\ghost{V_{3,1}}          &\qswap\qwx  &\qw             &\qswap\qwx}
  \end{align*}
  \caption{\label{fig:Pseudomodepicture} Qubit+ancilla embedding of the non-Markovian collision model, first three steps. 
    We have left out single qubit unitaries $U_0$ for the system qubit for simplicity.}
\end{figure}

The relevant Hilbert space for the bipartite state after $n$ collisions is $\hi_\sys\otimes\hi_{e,n}$. 
This Hilbert space is isomorphic to $\hi_\sys\otimes\hi_{e,1}$. We can 
use swap operators $S_{k,l}$ that swap the states between the environment particle 
$k$ and $l$ to implement this isomorphism 
\begin{align}\label{eq:37}
  W_{n}V_{n,n-1}=&(S_{n,1})^2W_n (S_{n,1})^2V_{n,n-1}\notag\\
  =&S_{n,1}W_1 S_{n,1} V_{n,n-1},
\end{align}
where the identity follows from  $S_{k,l}^{-1}=S_{k,l}=S_{k,l}^\dagger$.
The benefit of writing 
the collision model this way is that now the system qubit always interacts 
with the first environment particle which we will call {\it ancilla} from 
now on. Non-Markovian dynamics of the system qubit might  
be possible to embed into Markovian dynamics for the enlarged  
qubit+ancilla system. Circuit diagram for the collision model with ancilla qubit  
is presented in Fig.~\ref{fig:Pseudomodepicture}.
Clearly, the map $\Lambda$ can be written also as 
$\Lambda \kappa_n=
{\rm tr}_{n-1}\{S_{n,1}W_1S_{n,1}V_{n,n-1}\kappa_{n-1}(\kb{\varphi}{\varphi}\otimes\kb{0}{0})V_{n,n-1}^\dagger S_{n,1}W_1^\dagger S_{n,1}\}$
using the isomorphism between the original and qubit+ancilla collision models.
In the next subsection we illustrate the embedding construction with a specific example.
\subsection{Embedding collisional amplitude damping dynamics}\label{sec:coll-pseud-pict}
We take $W_i$ from Eq.~\eqref{eq:31} and $V_{i,i-1}$ from Eq.~\eqref{eq:2}. We assume
that initially the system qubit can be in an arbitrary pure state and ancilla qubit is 
in its ground state. 
Equation \eqref{eq:7} then gives the initial state $\kappa_1$, the 
explicit form for the initial state 
can be found in
Appendix \ref{sec:initial-condition}. 

From the initial condition and from the conservation of the excitation number 
we can deduce that  
$\bra{11}\kappa_n\ket{11}=\bra{11}\kappa_n\ket{10}=\bra{11}\kappa_n\ket{01}=
\bra{11}\kappa_n\ket{00}=\bra{00}\kappa_n\ket{11}=\bra{01}\kappa_n\ket{11}=\bra{10}\kappa_n\ket{01}=0$.
Thus the dynamics is confined in to a subspace $\{\ket{10},\ket{01},\ket{00}\}$ where the first index is for 
the system and the second for the ancilla.

The map $\Lambda$ is not necessarily completely positive as discussed earlier since
the state $\kappa_1$ is entangled. The reader may find 
the map $\Lambda$ written explicitly in the basis $\{\ket{10},\ket{01},\ket{00}\}$ in Appendix \ref{sec:map-lambda}.
However we know from the Kraus theorem 
that if we can write the map in the form $\Lambda\kappa = \sum_iB_i\kappa B_i^\dagger$,
where $\sum_iB_i^\dagger B_i=\id$ then the map is completely positive and 
trace preserving \cite{nielsen_quantum_2004}.
The search of the 
Kraus operators is non-trivial because of the correlated initial state~\cite{stelmachovic_dynamics_2001,stelmachovic_erratum:_2003}.
However, the fact that the Kraus operators can not be found using the standard approach
does not mean that they do not necessarily exist for this map \cite{tong_kraus_2006}.

Luckily, in this case we can find the Kraus decomposition by using the explicit construction of 
the map $\Lambda$. The map $\Lambda$ can be written as 
\begin{align}
  \kappa_{n}=\Lambda\kappa_{n-1} = \sum_{i=1}^2B_i\kappa_nB^\dagger_i,
\end{align}
where the operators $B_1$ and $B_2$ are  
\begin{align}
  B_1=&C\kb{00}{01},\label{eq:13}\\
  B_2=&c\kb{10}{10}-e^{i\phi}sS\kb{10}{01}\notag\\
  &-is\kb{01}{10}-ie^{i\phi}cS\kb{01}{01}+\kb{00}{00}\label{eq:14},
\end{align}
and where $s=\sin g$, $S=\sin G$, $c=\cos g$ and $C=\cos G$.

As we can see from Eqs.~\eqref{eq:13} and \eqref{eq:14} operator 
$B_2$ acts trivially on one dimensional subspace $\hi_g$ spanned by $\ket{00}$ and 
creates linear combinations in the subspace $\hi_e$ spanned by $\ket{10},\,\ket{01}$.
$B_1$ is just a transition operator from $\hi_e$ to $\hi_g$. Naming of these subspaces 
reflects the physical picture of a dissipative process where 
excitation resides in the two qubit system if the state lies in a subspace $\hi_e$ and 
the excitation has dissipated to the environment if the state of the 
two qubit system is in $\hi_g$. Naturally, since we are dealing with quantum systems the 
state could be a superposition state of vectors in $\hi_g$ and $\hi_e$. 

Indeed, the initial state $\kappa_1$ is of the superposition form.
Namely, 
$\kappa_1=\kb{\psi}{\psi}$, where 
$\ket{\psi} = c_0\ket{00}+c1(\cos g\ket{01}-i\sin g\ket{10})$, see Appendix \ref{sec:initial-condition}. 
For future reference 
we write $\ket{\tilde\psi_1}=g_1\ket{00}+\alpha_1\ket{01}+\beta_1\ket{10}\equiv\ket{\psi}$.
Symbol $\ket{\tilde\varphi}$ is used to denote that vector $\varphi$  
is not necessarily normalized to unity.

Starting from a state $\kappa_1$ the dynamical map $\Lambda$ produces a mixed state 
$\kappa_2 = \sum_i B_i\kappa_1 B_i^\dagger$, where $B_1\kappa_1B_1^\dagger$  can be 
written as $|\xi_2|^2\kb{00}{00}$. 
The other component of the mixed state $\kappa_2$ is 
$B_2\kappa_1B_2^\dagger$ which can be written as $\kb{\tilde\psi_2}{\tilde\psi_2}$, where 
$\ket{\tilde\psi_2} = g_2\ket{00}+\alpha_2\ket{10}+\beta_2\ket{01}$. Continuing this 
way we deduce that for arbitrary $n$ we can write the state $\kappa_n$ as 
\begin{align}\label{eq:11}
  \kappa_n = \kb{\tilde\psi_n}{\tilde\psi_n}+|\xi_n|^2\kb{00}{00},
\end{align}
with $|\xi_1|^2=0$ and $\ket{\tilde\psi_1}=\ket{\psi}$.

By comparing the coefficients of states $\kappa_{n+1}$ and $\kappa_n$ 
we see that the parameters $\alpha_n,\,\beta_n$ and $\xi_n$ satisfy the 
following recursion relation
\begin{align}
  \alpha_{n+1}=&(\alpha_n c-\beta_ne^{i\phi}sS),\label{eq:20}\\
  \beta_{n+1}=&-(\alpha_nis+\beta_ne^{i\phi}sS),\label{eq:28}\\
  \xi_{n+1}=&C\beta_n+\xi_n\label{eq:29},
\end{align}
with initial conditions $\xi_1=c_0$, $\alpha_1=c_1\cos g$ and 
$\beta_1 = -ic_1\sin g$.

By tracing out the environment qubit from $\kappa_n$ and 
by using the amplitude damping channel structure of the map $\Phi_n$ we can write the 
function $\eta_n$ in terms of collision model as  
$\eta_n=\alpha_n/(c_1c_0^*)$. 
Then from the recursion relations \eqref{eq:20},\eqref{eq:28} we obtain
Eq.~\eqref{eq:32} for the function $\eta_n$.

\subsubsection{Continuous limit of $\Lambda_n$}\label{sec:cont-limit-lambd}
Note that in this section we will use physical units, eg. $(t,\Gamma,\Omega,\lambda)$, since the discussion in
this section becomes most natural by this choice.
Map $\Lambda$ describes discrete Markovian dynamics. 
We expect that the continuous limit is given by a dynamical semigroup
$e^{\Li t}$. 
For a small time interval $\delta t$ we can use the 
dynamical semigroup to write  
$\kappa_{n+1}\approx \kappa_n+\delta t\Li\kappa_n$. 
Then we demand that to first order in $\delta t$
we have
\begin{align}\label{eq:17}
  (\id+\delta t\Li)\kappa_n = &\Lambda\kappa_n.
\end{align}

It turns out that the continuous limit of $\Lambda$ is given by the following 
Lindblad master equation
\begin{align}\label{eq:16}
  \Li\kappa =& -i \big(\left[
               \alpha\sigma_+\sigma_-\otimes\id+\beta\id\otimes\sigma_+\sigma_-,\kappa\right]\notag\\
             &+\delta\left[\sigma_+\otimes\sigma_-+\sigma_-\otimes\sigma_+,\kappa\right]\big)\notag\\
             &-\gamma\left(\{\id\otimes\sigma_+\sigma_-,\kappa\}+2\id\otimes\sigma_-\kappa\id\otimes\sigma_+\right),
\end{align}
where $\alpha,\,\beta,\,\delta$ and $\gamma$ are free parameters.

By choosing the $\delta t$ dependence of the collision model parameters as in 
Eqs.~\eqref{eq:22}-\eqref{eq:30}
and then by expanding by expanding the state  $\Lambda\kappa$ to first order in $\delta t$ and 
comparing to $(\id+\delta t\Li)\kappa$ we 
obtain the following conditions for the free parameters:
$\alpha = 0$, $\beta = \Omega$, $\delta =\sqrt{\frac{\lambda\Gamma}{2}}$ and 
$\gamma = \Gamma$. 
What we have obtained is that in the continuous limit 
the generator of the two qubit dynamics is given by
\begin{align}\label{eq:21}
\Li\kappa =& -i \big(\Omega\left[\id\otimes\sigma_+\sigma_-,\kappa\right]\notag\\
               &+\sqrt{\frac{\lambda\Gamma}{2}}\left[\sigma_+\otimes\sigma_-+
                 \sigma_-\otimes\sigma_+,\kappa\right]\big)
                 \notag\\
               &-\Gamma\left(\{\id\otimes\sigma_+\sigma_-,\kappa\}+2\id\otimes\sigma_-
                 \kappa\id\otimes\sigma_+\right).
\end{align}
We can identify this as the pseudomode master equation for the spontaneously decaying two level system 
when the environment is initially in the zero temperature vacuum state \cite{garraway_nonperturbative_1997}.

\section{Conclusions}\label{sec:conclusions}
In this work we have studied the possibility to use collisional models for studying 
indivisible quantum dynamics and non-Markovianity. We proposed in Sec.~\ref{sec:coll-model-indiv} a collisional
model that shows potential for constructing indivisible dynamical maps.  

We also showed in Sec.~\ref{sec:discrete-maps} that 
our model can be used to simulate the process of spontaneous decay of a two level system to a structured reservoir, 
which leads to amplitude damping,
exactly. We discussed in Sec.~\ref{sec:non-mark-open} the non-Markovianity of the collision model dynamics which is 
discrete in time and its relation to the indivisibility of the time-continuous dynamical map. 

In Sec.~\ref{sec:coll-pseud-pict} we showed how one can distinguish a particular environment particle 
and interpret it as a pseudomode. We also showed how this leads to positively divisible dynamics for the 
combined bipartite system. We studied in detail the case of amplitude damping and found out that in the time 
continuous limit we obtain the well known pseudomode master equation.

We can conclude that the collision model approach shows great potential for studying 
quantum memory effects in open quantum system dynamics. They might open a way to 
study continuous monitoring of the open system in the case where the dynamical map is 
indivisible. Clearly, a very wide open question is what properties are generally required 
for the operators $W_i$ and $V_{i,i-1}$ so that the collision model dynamics is indeed indivisible.

\begin{acknowledgments}
  The authors acknowledge fruitful discussions with Richard Hartmann and Alexander Eisfeld.
\end{acknowledgments}
\appendix
\section{Proof of Eqs.~\eqref{eq:22}-\eqref{eq:30}}\label{sec:proof-eqs.-eqref}
Here we show how the continuous-in-time limit for the collisional model dynamics given 
by Eqs.~\eqref{eq:33}\eqref{eq:31}\eqref{eq:2} is obtained.
We begin by assuming that $g\propto \delta \tau $. Then we 
assume that $\delta \tau \ll 1$ and in Eq.~\eqref{eq:32} we 
use $\cos^2 g\approx 1$, $\tan^2 g\approx g^2$. 
With these the difference
$\Delta \eta_n = \eta_n -\eta_{n-1}$ can be written as 
$\Delta \eta_n = -g^2\sum_{j=0}^{n-2} \eta_j  (-ie^{i\phi}\cos g \sin G)^{n-1-j}$.
When we substitute to the above equation  $\cos g\approx 1$ and Eqs.~\eqref{eq:22}-\eqref{eq:30} 
we obtain $\Delta \eta_n = -\lambda\frac{\tlambda}{2}\delta \tau\sum_{j=0}^{n-2}\delta\tau\eta_j(e^{-(1+i\tOmega)\delta\tau})^{n-1-j}$.    
Next we take the continuous limit $\delta \tau\to 0$, where we are allowed to replace $(n-2)\delta\tau \approx (n-1)\delta\tau\approx \tau$,
and we obtain
\begin{align}\label{eq:42}
  \dot{\eta_\tau}=\lim_{\delta \tau\to 0}\frac{\Delta \eta_n}{\delta\tau}=-\frac{\tlambda}{2}\int_{0}^\tau
  \meas \tau'\,e^{-(1+i\tOmega)(\tau-\tau')}\eta_{\tau'},
\end{align}
which is the defining equation for $\eta_{\tau}$.
\section{Initial condition}\label{sec:initial-condition}
The initial state for the open system+ancilla evolution in Sec.~\ref{sec:coll-pseud-pict} is 
given by $\kappa_1=W_1\rho_0\otimes\kb{0_1}{0_1}W_1^\dagger$ which can be written 
explicitly as  
\begin{align}\label{eq:8}
  \kappa_1=&\rho_{11}\left(\cos^2g\kb{10}{10}+\sin^2g\kb{01}{01}\right)+\rho_{00}\kb{00}{00}\notag\\
           &+i\rho_{11}\cos g\sin g\left(\kb{10}{01}-\kb{01}{10}\right)\notag\\
           &+\rho_{10}\left(\cos g\kb{10}{00}-i\sin g \kb{01}{00}\right)\notag\\
           &+\rho_{01}\left(\cos g\kb{00}{10}+i\sin g\kb{00}{01}\right).
\end{align}
$\kappa_1$ can be written in terms of vectors 
$\ket{\tilde\varphi_1}=c_1(\cos g\ket{10}-i\sin g\ket{01})$, $\ket{\tilde\varphi_2}=c_0\ket{00}$,
where $|c_0|^2=\rho_{00}$, $|c_1|^2=\rho_{11}$, $c_0c_1^*=\rho_{01}$ and $c_1c_0^*=\rho_{10}$ as 
$\kappa_1=\kb{\tilde\varphi_1}{\tilde\varphi_1}+\kb{\tilde\varphi_2}{\tilde\varphi_2}+
\kb{\tilde\varphi_1}{\tilde\varphi_2}+\kb{\tilde\varphi_2}{\tilde\varphi_1}$.

\section{Map $\kappa_{n+1}=\Lambda\kappa_n$}\label{sec:map-lambda}
Here we write explicitly the dynamical map studied in Sec.~\ref{sec:coll-pseud-pict}.
The map $\Lambda:\kappa_n\mapsto \kappa_{n+1}$ can be written in the basis $\{\ket{10},\ket{01},\ket{00}\}$ as 
\begin{widetext}
  \begin{align}\label{eq:12}
    \kappa_{n+1}=&\left(\begin{smallmatrix}
      \kappa_{10,10} c^2+ s^2 S^2 \kappa_{01,01}-csS\left(e^{-i \phi }\kappa_{10,01}+
          e^{i \phi } \kappa_{01,10}\right) 
      & i \left(c s \kappa_{10,10}+S c^2 e^{-i \phi } \kappa_{10,01}
          -e^{i \phi }S s^2 \kappa_{01,10}+csS^2 \kappa_{01,01}\right) 
      & c \kappa_{10,00}-e^{i \phi } s S \kappa_{01,00} \\
      -i \left(c s \kappa_{10,10}+S c^2 e^{i \phi }\kappa_{01,10}-csS^2 \kappa_{01,01}
          -e^{-i \phi } s^2 \kappa_{10,01}\right) 
      & \kappa_{10,10} s^2+c s S \left(e^{-i \phi }  \kappa_{10,01}
        +e^{i \phi }  \kappa_{01,10}\right)+c^2 S^2 \kappa_{01,01} & -i \left(s \kappa_{10,00}+e^{i \phi } cS 
        \kappa_{01,00}\right) \\
      c \kappa_{00,10}-e^{-i \phi } s S \kappa_{00,01} 
      & i( s \kappa_{00,10}+c S e^{-i\phi} \kappa_{00,01}) 
      & C^2\kappa_{01,01} +\kappa_{00,00}\\
    \end{smallmatrix}\right),
  \end{align}
\end{widetext}
where $c=\cos g$, $s=\sin g$, $S=\sin G$ and $C=\cos G$ and 
$\kappa_{ij,kl}=\bra{ij}\kappa_n\ket{kl}$. Validity of the Kraus operators $B_1,B_2$ 
can be easily verified using the above explicit expression.
\bibliography{QubitCollisionManu}
\end{document}